\documentclass{article} 
\usepackage{preprint,times}

\usepackage{amsmath}
\usepackage{amssymb}
\usepackage{amsthm}
\usepackage{hyperref}
\usepackage{url}
\usepackage{algorithm}
\usepackage{algpseudocode}
\usepackage{graphicx}
\usepackage{natbib}  
\usepackage{caption} 
\usepackage{booktabs}
\usepackage{multirow}
\usepackage{makecell}
\usepackage{xcolor}

\usepackage{amsmath,amsfonts,bm}

\def\eqref#1{equation~\ref{#1}}

\def\plaineqref#1{\ref{#1}}

\def\1{\bm{1}}

\def\vc{{\bm{c}}}

\def\vn{{\bm{n}}}

\def\vp{{\bm{p}}}

\def\vr{{\bm{r}}}

\def\vv{{\bm{v}}}
\def\vw{{\bm{w}}}
\def\vx{{\bm{x}}}
\def\vy{{\bm{y}}}

\def\mSigma{{\bm{\Sigma}}}

\DeclareMathAlphabet{\mathsfit}{\encodingdefault}{\sfdefault}{m}{sl}
\SetMathAlphabet{\mathsfit}{bold}{\encodingdefault}{\sfdefault}{bx}{n}

\newcommand{\E}{\mathbb{E}}

\newcommand{\R}{\mathbb{R}}

\newcommand{\vE}{{\bm{E}}}
\newcommand{\vX}{{\bm{X}}}
\newcommand{\vJ}{{\bm{J}}}
\newcommand{\vI}{{\bm{I}}}
\newcommand{\bDelta}{{\bm{\Delta}}}
\newcommand{\vB}{{\bm{B}}}
\newcommand{\vC}{{\bm{C}}}
\newcommand{\vM}{{\bm{M}}}
\newcommand{\vY}{{\bm{Y}}}
\newcommand{\vR}{{\bm{R}}}
\newcommand{\vG}{{\bm{G}}}
\newcommand{\vS}{{\bm{S}}}
\newcommand{\valpha}{{\bm{\alpha}}}
\newtheorem{proposition}{Proposition}

\newcommand{\methodname}{SCAMP}

\title{SCAMP: Sparse-Anchor Control Is One Small Projection}

\author{Pengcheng Fang\thanks{Equal contribution.}$^{,1,2}$,
Tengjiao Sun\footnotemark[1]$^{,1,2}$,
Xiaoyu Zhan$^{2,3}$,
Yanwen Guo$^{3}$,
Hansung Kim$^{1}$,
Xiaohao Cai$^{1}$,
Dongjie Fu\thanks{Corresponding author.}$^{,2}$\\
$^{1}$University of Southampton \quad
$^{2}$Mogo AI Ltd. \quad
$^{3}$Nanjing University}

\ppfinalcopy

\begin{document}
\maketitle
\lhead{}

\begin{abstract}
Text fixes what a motion is, but not where the body must be. Authoring needs
sparse anchors: chosen joints, at chosen frames, at given positions. Meeting
them currently costs a conditioning branch trained for the task, or hundreds
of per-clip optimisation steps in the architecture's native variables. In any
generator that decodes a continuous state through a frozen differentiable
decoder, the anchors ask for little and leave most of the state free: a few
hundred numbers against tens of thousands, so the states that meet them form
a set within a few hundred dimensions of the state space itself. Every control
method is a choice inside that set, and the choices differ along the
directions the anchors cannot see and the motion can. Our solver,
\methodname{} (sparse-anchor control as a minimum-norm projection), makes the
choice that moves none of them: damped Gauss--Newton in the space of the
anchors, through the frozen decoder alone, training-free, one dimensionless
damping constant. Every increment is a combination of the rows of the anchors'
Jacobian, so the correction is orthogonal to everything they never see, and
the system solved is the request's size and not the state's. Applied unchanged
to seven published generators spanning diffusion, token and latent designs, in
anchor error it matches or exceeds every released control method it is
measured against, and closes the anchors on the hosts that ship none. Held to
those rows, a correction can only take the shapes the decoder admits, so its
cost is the decoder's, and a fixed solver makes that cost measurable: every
generator pays in foot skating, and how much divides by decoder family. A
decoder that spreads a change of state over a window keeps its mean
contact-speed skating within a small multiple of the unconstrained
generator's own; an analytic recovery displaces the body itself, and skating
multiplies several times over. Our
generator is built to that criterion: a windowed decoder and a sampler trained
through the correction, so it meets its anchors within a shallow solve at
sub-second cost. The decoder's temporal support is a design criterion for
controllable motion generation.
\end{abstract}

\section{Introduction}
\label{sec:intro}

A foot on a mark, a hand on a door handle, a root passing through a
waypoint: the moment a text-to-motion model meets an authoring task,
the specification stops being a sentence and becomes a small set of
Cartesian constraints on particular joints at particular frames ---
\emph{sparse anchors}. Generators meet anchors today with machinery
bought at real expense: a conditioning branch trained for the task,
or per-clip optimisation cast in the architecture's native variables
--- guidance on the
sample~\citep{karunratanakul2023gmd,xie2024omnicontrol}, descent on
the initial noise~\citep{karunratanakul2024dno}, re-imposition of
observed frames during denoising~\citep{cohan2024condmdi}, descent in
a learned latent space~\citep{dai2024motionlcm}. The machinery is out of
proportion to the thing specified: a few hundred scalar constraints,
set against a state of $10^4$--$10^5$ dimensions.

The disproportion has a mathematical root. Every one of these mechanisms runs on a generator that produces motion by decoding a continuous state $\vE$ --- a diffusion sample, a latent, a token model's code embedding --- through a frozen differentiable decoder; call it a \emph{host}. Under that one shared
assumption, anchor control is a single problem: find the state
nearest the host's own output whose decoded motion meets the anchors,
\begin{equation}
  \min_{\vE}\ \tfrac12\,\lVert \vE-\vE_0\rVert^2
  \quad\text{subject to}\quad \vr(\vE)=\bm{0},
  \label{eq:intro_problem}
\end{equation}
where $\vE_0$ is the host's own sample and $\vr$ stacks the anchor
deviations of the decoded motion.

What that problem asks is small; what it leaves open is not. Write $\vJ$ for
the Jacobian of $\vr$: one row per anchor coordinate, $m$ of them, against one
column per coordinate of the state, $P$ of those --- a few hundred against
tens of thousands. The states that meet the anchors therefore form a set of
dimension $P-m$, and every control method, a trained branch or a guidance
path or re-imposed frames or a latent descent, is a choice inside it. The
choices are not interchangeable: the anchors are blind to those directions and
the motion is not. \methodname{} takes the one that moves none of them.
Each accepted increment is a combination of the rows of $\vJ$, so it is
orthogonal to $\ker\vJ$, and the correction as a whole is orthogonal to every direction that
stayed invisible to the anchors along the way --- five accepted steps enter at
most $5m$ of the state's directions and leave the rest where the generator put
them.

Making that choice costs one linear solve. The objective of
problem~\plaineqref{eq:intro_problem} is exactly quadratic and only the constraint
is linearised, which gives a step of the residual's own dimension, taken
through the decoder rather than the sampler. The solver, \methodname{}, is
damped Gauss--Newton with a single damping constant made dimensionless by the
Jacobian's own scale, and the system it solves is $m\times m$ on every host:
the request sets its size, not the state. Applied unchanged to seven published hosts it matches or exceeds each released control method in anchor error, without training and, on MDM, MaskControl and OmniControl, in a fraction of its time (\S\ref{sec:crossmodel}).

Because the correction lives in those rows, the shapes it can take are the
decoder's and so is what they cost. Every representation
admits the correction, and every one pays for it in foot skating on the way.
The size of the correction fails to predict the cost: the host that
moves the motion farthest pays a fraction of what the host that moves it least pays. What
tracks it is the decoder's temporal support, the number of frames
that feel a change of one state coordinate. A decoder that reconstructs each frame from a window of state answers an anchor with a local deformation; an analytic recovery moves the body itself, either within the anchored frames or, when the root is integrated, as a velocity bias across the whole prefix. Grouped by decoder family the hosts separate without overlap,
and an intervention inside a single host moves the cost in whichever
direction its decoder leaves open: widening the update's temporal
basis lowers it on a frame-local decoder and buys nothing on a
windowed one (\S\ref{sec:law}).

The finding motivates a windowed decoder. We keep one and build the sampler around it: a
continuous-time Markov chain over motion tokens, conditioned on the
anchor residual and trained through the terminal correction
(\S\ref{sec:model}). It meets its anchors at sub-second cost.

\noindent\textbf{Contributions.} \textbf{(i)~One operator for anchor
control.} Damped Gauss--Newton in the space of the anchors, through the frozen
decoder alone, taking the increment that moves nothing the anchors cannot see
and solving $m\times m$ whatever the size of the state --- training-free, one
dimensionless damping constant, tuned per host nowhere. Retrofitted unchanged
onto seven published generators, it matches or exceeds in anchor error the
released control method of every host that ships one, and closes the anchors
on the two that ship none
(\S\ref{sec:crossmodel}). \textbf{(ii)~The cost of a correction, read as a
property of the representation.} One fixed operator makes the generator the
experimental variable, and the physical cost divides by decoder family, with
the decoder's temporal support a causal factor in it: the families separate
without overlap in jerk and in mean contact speed (\S\ref{sec:law}). \textbf{(iii)~A generator built to that criterion.}
A codebook-geometric continuous-time Markov chain over motion tokens,
keeping a windowed decoder and trained through the correction itself,
reaching $0.083$\,m anchor error at FID $0.102$ in $0.50$\,s per clip
(\S\ref{sec:ours}).

\section{Related work}
\label{sec:related}

\noindent\textbf{Control native to motion generators.}
GMD~\citep{karunratanakul2023gmd} and
OmniControl~\citep{xie2024omnicontrol} add a classifier-style gradient
to the denoising steps of a raw-motion diffusion model; the gradient
is evaluated through the decoder's map to joint positions, which is
\eqref{eq:gn_update} up to a step size in its large-$\mu$ limit.
DNO~\citep{karunratanakul2024dno} and
MotionLCM's CLT~\citep{dai2024motionlcm} instead treat the initial
noise or the latent as the decision variable and run hundreds of Adam
steps through the whole sampler; our correction backpropagates
through the decoder alone, never through the sampler. On the trained
side, CondMDI~\citep{cohan2024condmdi} learns a keyframe-conditioned
model and imputes observed frames during sampling, and
MaskControl~\citep{pinyoanuntapong2025maskcontrol} trains a control
branch on a masked token model with token-space logit optimisation on
top --- the strongest prior method in our comparison. Imputation is
our undamped limit: at $\mu\to0$ with a time-selection Jacobian,
whose rows are orthonormal, \eqref{eq:gn_update} is the exact
projection onto the anchors --- the most concentrated correction
available, and \S\ref{sec:law} measures what that concentration
costs.

\noindent\textbf{Inverse problems inside a frozen prior.}
Outside motion this is a developed literature.
RePaint~\citep{lugmayr2022repaint} resamples observed regions;
manifold-constrained gradients~\citep{chung2022mcg}, posterior
sampling~\citep{chung2023dps} and pseudo-inverse
guidance~\citep{song2023pigdm} steer each denoising step toward the
measurement; the null-space decomposition~\citep{wang2023ddnm}
projects each iterate onto the measurement set --- for a linear
decoder, the $\mu\!\to\!0$ limit of our own update;
universal~\citep{bansal2024universal} and energy~\citep{yu2023freedom}
guidance backpropagate an arbitrary objective through the denoiser.
Our operator belongs to this family, carried from gradient steps to
damped Gauss--Newton steps and to a general host state. We use it as
an \emph{instrument}: one solver unchanged across hosts makes the
representation the variable (\S\ref{sec:law}). Differentiable-optimisation
layers~\citep{amos2017optnet,pineda2022theseus} supply the machinery
for training through such a solve, as \S\ref{sec:model} does.

\noindent\textbf{Classical constrained motion editing.}
Spacetime constraints~\citep{witkin1988spacetime,gleicher1997motion}
and style-based IK~\citep{grochow2004styleik} solve the same
projection, but against hand-designed regularisers rather than a
learned prior. Here the regulariser is the host's own decoder, whose
temporal coupling applies to every iterate, and \S\ref{sec:law}
measures how far it reaches.

\section{Method}
\label{sec:method}

\begin{figure}[t]
\centering
\includegraphics[width=0.76\textwidth]{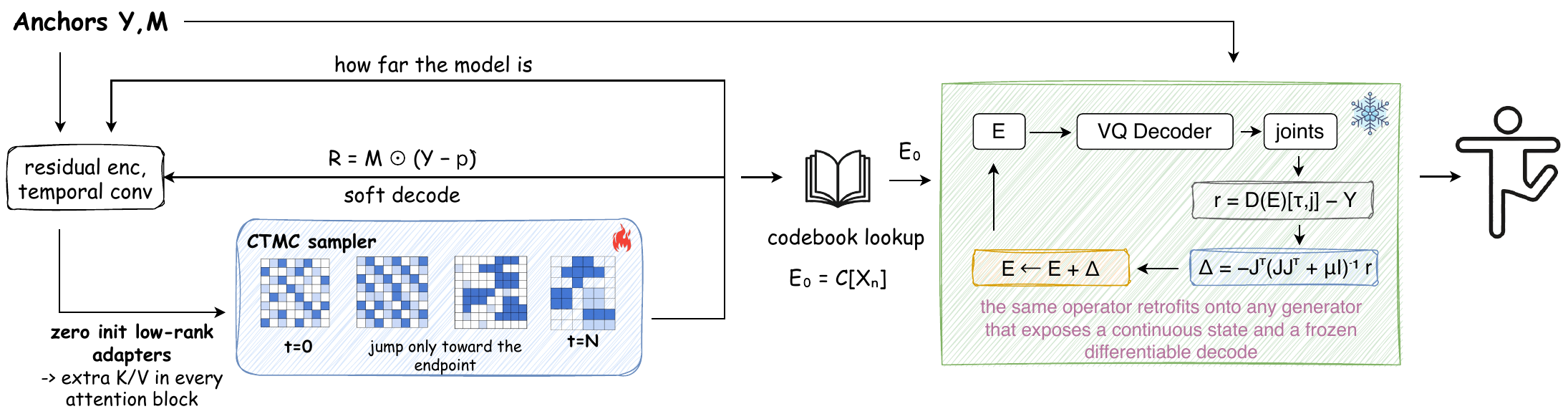}
\caption{The anchors enter twice: during sampling as a feedback signal, the
current clean posterior soft-decoded to joints $\hat{\vp}$ and the model
conditioned on $\vR=\vM\odot(\vY-\hat{\vp})$ through extra keys and values;
at the end as the target of one damped Gauss--Newton solve through the frozen
decoder. The tokeniser is frozen throughout and its decoder is windowed, so a
correction spreads over its receptive field.}
\label{fig:main}
\end{figure}

\subsection{The problem, and what a host must expose}
\label{sec:problem}

The layer consumes two objects: a continuous state $\vE\in\R^{P}$ the
generator produces, and a frozen differentiable map $\mathcal D$ from
that state to Cartesian joint positions. Any generator supplying them
is a \emph{host} --- $\vE$ is a diffusion sample, a latent, or a token
model's continuous code embedding, and $\mathcal D$ composes the
host's learned decoder, where it has one, with the analytic recovery
of joint positions from its motion representation. On token hosts we decode the corrected state directly, so the correction reaches the motion in full rather than being rounded back onto the codebook.

A control request is a set of $A$ anchors $\{(\tau_a,j_a,\vy_a)\}$:
joint $j_a$ at position $\vy_a$ in frame $\tau_a$. Stacking the
deviations gives the residual
\begin{equation}
    \vr(\vE)
    =
    \left[\,
        \mathcal D(\vE)[\tau_a,j_a] - \vy_a
    \,\right]_{a=1}^{A}
    \in\R^{m},
    \qquad
    m = 3A ,
    \label{eq:residual}
\end{equation}
and the request specifies problem~\plaineqref{eq:intro_problem}: the state nearest the host's own output that zeros $\vr$. Anchors are sparse, so $m$ is orders of
magnitude below $P$.

\subsection{The damped Gauss--Newton step}
\label{sec:gn}

Write $\vJ=\partial\vr/\partial\vE\in\R^{m\times P}$. Linearising the
constraint of problem~\plaineqref{eq:intro_problem} at $\vE_0$ gives its
first step: the smallest increment that meets the linearised anchors,
\begin{equation}
    \min_{\bDelta}\ \tfrac12\lVert\bDelta\rVert_2^2
    \quad\text{subject to}\quad
    \vJ\bDelta = -\vr .
    \label{eq:gn_objective}
\end{equation}
Stationarity gives $\bDelta=\vJ^{\top}\vv$ with
$\vJ\vJ^{\top}\vv=-\vr$: the increment lies in the row space of $\vJ$,
and the unknown is the multiplier $\vv$, one coordinate per anchor
coordinate. That row space is the whole of what the anchors can reach: $\vJ$
has $m$ rows against $P$ columns, so its kernel has dimension $P-m$, and the
increment is orthogonal to all of it, $\langle\bDelta,\vn\rangle =
\vv^{\top}\vJ\vn = 0$ for every $\vn\in\ker\vJ$. Confining the increment to a trust region, equivalently
relaxing the equality to a quadratic penalty, adds a multiple of the
identity to that system~\citep{more1978lm},
\begin{equation}
    \boxed{\;
    \bDelta
    =
    -\,\vJ^{\top}
    \bigl(\vJ\vJ^{\top}+\mu\vI_m\bigr)^{-1}
    \vr
    \;}
    \label{eq:gn_update}
\end{equation}
which leaves the increment in the row space and so leaves the orthogonality
where it was, and keeps the linear system at $m\times m$: the request sets its
size, not the state, $m{=}144$ at the anchor density of \S\ref{sec:exp}
against $P$ in the tens of thousands. The matrix inverted,
$\vJ\vJ^{\top}+\mu\vI_m$, is positive semidefinite by construction and
positive definite for every $\mu>0$, so a step exists for any request on any
host. The same increment written in the state,
$-(\vJ^{\top}\vJ+\mu\vI_P)^{-1}\vJ^{\top}\vr$, asks for a $P\times P$ inverse;
only one of the two is computable at this size. A
correction of $R$ accepted steps therefore enters at most $Rm$ of the state's
directions and leaves the rest of the state exactly where the host put it.
The iteration repeats this step from the corrected state and returns the endpoint of that path; how far it moves the decoded motion is measured per host (Displ.\ in Table~\ref{tab:families}). At $\mu\to0$ and full row rank each step
is the minimum-norm increment of \eqref{eq:gn_objective} --- the
property \S\ref{sec:coupling} uses --- and applied to Cartesian joint
targets \eqref{eq:gn_update} is the damped least-squares step of
singularity-robust inverse
kinematics~\citep{wampler1986dls,nakamura1986singularity,chiaverini1994dls},
whose undamped min-norm form goes back
to~\citet{benisrael1966newton}. What differs from that setting is the shape of
$\vJ$. In inverse kinematics it is analytic and the unknowns are a few dozen
joint angles, so the system can be square or overdetermined and the damping
carries it through kinematic singularities. Here it is read from the frozen
decoder by $m$ reverse-mode products, and sparse anchors against a generator's
state make $m\ll P$ by construction: the system is always the underdetermined
one, and the damping is the trust region the linearisation asks for.

\noindent\textbf{The loop.} We iterate: build $\vJ$, solve
\eqref{eq:gn_update}, and accept the trial only if it strictly
decreases the merit $\Phi(\vE)=\lVert\vr(\vE)\rVert_2^2$. A rejected
trial raises the damping and re-solves the same system; acceptance
lowers it. When every trial at a state is rejected, the solve returns
the last accepted state (constants in Appendix~\ref{app:solver}). As
$\mu$ grows the step approaches $-\vJ^{\top}\vr/\mu$, a descent
direction for $\Phi$, so a large enough damping yields an accepted
step at any non-stationary state.

Solve depth --- the number of accepted steps --- is chosen per
request at inference and sets the precision.

\subsection{Scale-normalised damping}
\label{sec:damping}

One dimensionless constant has to serve hosts whose states differ in
scale by orders of magnitude: a code embedding, a VAE latent and a
normalised motion feature give Gram matrices whose diagonals are not
comparable. We set
\begin{equation}
    \mu
    \;=\;
    \lambda\cdot
    \frac{1}{m}\operatorname{tr}\!\left(\vJ\vJ^{\top}\right),
    \label{eq:scale_damping}
\end{equation}
which makes $\lambda$ dimensionless: under $\vE\mapsto\alpha\vE$ we
have $\vJ\mapsto\vJ/\alpha$, so $\mu\mapsto\mu/\alpha^{2}$ and
$\bDelta\mapsto\alpha\bDelta$ --- the iterate is exactly equivariant,
and rescaling $\vr$ leaves it unchanged. A single dimensionless $\lambda_0=10^{-2}$, set once, initialises the
loop on every host in this paper.

Scalar damping is
Levenberg's~\citep{levenberg1944method}; the accept-and-retry
schedule and the curvature-derived scale are
Marquardt's~\citep{marquardt1963algorithm}, whose scaling is
per-coordinate where \eqref{eq:scale_damping} is a single scalar;
the trust-region reading and adaptive scaling are
\citet{more1978lm}'s.

\subsection{Metric and reparameterisation}
\label{sec:metric}

Solving \eqref{eq:gn_objective} for a reparameterised increment
$\bDelta=\vB\valpha$ of smallest $\lVert\valpha\rVert$ gives
$\bDelta=-\mSigma\vJ^{\top}(\vJ\mSigma\vJ^{\top}+\mu\vI)^{-1}\vr$ with
$\mSigma=\vB\vB^{\top}\succeq0$, a preconditioner: a rectangular $\vB$
confines the increment to the subspace its columns span. The damping
scale follows it,
$\mu=\lambda\operatorname{tr}(\vJ\mSigma\vJ^{\top})/m$, so a change of
$\vB$ leaves the effective regularisation where it was. We take $\vB=\vI$, and so $\mSigma=\vI$, on the evidence of
Appendix~\ref{app:metric}, except in the intervention of
\S\ref{sec:law}, where the
columns of $\vB$ are temporally smooth atoms of controlled width and
widening them restricts the increment to smoother shapes.

\subsection{What the state costs: temporal support}
\label{sec:coupling}

For every damping the increment lies in the row space of $\vJ$. Write
$\vJ=\vS\vG$ for the decoder's full Jacobian $\vG$ and the selection
$\vS$ that reads the anchored coordinates: the decoded displacement is
then a combination of the columns of $\vG\vG^{\top}\vS^{\top}$, rows
of the decoder's own Gram, which for a convolutional decoder is the
autocorrelation of its temporal response and so reaches about twice as
far. The host fixes the family of shapes a correction can take; the
request selects which of them, and how much.

This width is a property of the frozen map $\mathcal D$ alone, and is measured
without a solve: perturb one coordinate of the state at one frame,
decode, and read how far in time the displacement extends. We report
it as $w_{\text{eff}}=\sum_t d_t/\max_t d_t$ for $d_t$ the joint
displacement at frame $t$: one for a strictly frame-local decode, growing with the spread. Coordinates are drawn uniformly and the median reported, so it reports the reach of a typical coordinate; the few integrated root channels of an analytic recovery reach further, and \S\ref{sec:law} reads them in the displacement they produce.

The families differ in kind. A learned decoder of temporal receptive
field $w$ answers an anchor with a deformation supported on a window
--- a shape the prior already produces, since the decoder was fitted
to plausible motion. An analytic recovery has no such window. From
absolute-root features it is frame-local in every channel, so the
increment stays inside the anchored frames. From relative-root
features the root is integrated, so the row of an anchor at frame
$\tau$ is constant over the whole preceding interval and the
minimum-norm increment is a velocity bias of size
$\lVert\vr\rVert/\tau$ applied to every frame up to $\tau$: the body
is displaced, gently but continuously, for the whole prefix.

The cost is measured as foot skating, the standard
physical-plausibility artefact of motion synthesis: a foot within
$5$\,cm of the ground that moves horizontally faster than $0.5$\,m/s,
over a five-frame window at $20$\,fps, counts against the
clip~\citep{xie2024omnicontrol}. Two things decide the count, the speed each frame carries and
how many contact frames carry it, and the three correction shapes sit
differently on both. A windowed deformation is compact and spread over the decoder's
window, so the speed it adds at any one contact frame is small. A frame-local correction puts a whole displacement between
two frames, far above the threshold. A root bias is small per frame
and applies to the entire body, so every foot planted anywhere in the
prefix is dragged with it.

\subsection{A generator built to the criterion}
\label{sec:model}

We address the shape by keeping a frozen VQ tokeniser of the MoMask
family~\citep{guo2024momask}, whose convolutional decoder is a windowed decode, so a correction is spread over its receptive field; ours uses a single quantiser at unit temporal stride, so
one token corresponds to one frame. We address the magnitude by
conditioning the sampler on the anchor residual and training it
\emph{through} the terminal solve. The
sampler is a continuous-time Markov chain over the token sequence,
whose rates admit only jumps toward the sampled endpoint: an
intermediate sequence stays on the codebook's local geometry and
decodes to a nearby motion, so the anchor residual is informative from the
first steps and the training gradient reaches the emitted state through it.

\noindent\textbf{Codebook-geometric CTMC.} We formulate generation as
continuous-time discrete diffusion over the token
sequence~\citep{austin2021d3pm,campbell2022ctmc} and instantiate its
endpoint-conditioned path~\citep{campbell2024discreteflow,gat2024discreteflow}
with the geometry of the frozen token embedding table. We adopt the prototype distance of \citet{fang2026mogeflow}: for normalised prototypes $\bar{\vc}_v$, $\delta(i,j)=\lVert\bar{\vc}_i-\bar{\vc}_j\rVert_2^{2p}$, whose exponent $p$ sets how sharply the rates concentrate on near neighbours. With an increasing
schedule $\beta(t)=c_{\beta}(t/(1-t))^{\alpha}$,
$c_\beta,\alpha>0$, the endpoint-conditioned path
\begin{equation}
  q_t(k\mid y)\;\propto\;\exp\!\bigl[-\beta(t)\,\delta(k,y)\bigr]
  \label{eq:path}
\end{equation}
starts uniform and, for pairwise distinct prototypes, concentrates on the
endpoint $y$ as $t\to1$. The
off-diagonal rates
\begin{equation}
  Q^{y}_t(i,k)\;=\;q_t(k\mid y)\,\dot\beta(t)\,
  \bigl[\delta(i,y)-\delta(k,y)\bigr]_{+},\qquad k\neq i,
  \label{eq:endpoint_gen}
\end{equation}
admit only jumps to codes strictly closer to $y$, so a position is
absorbed once it reaches its endpoint and is released only when the
endpoint sample changes.

\begin{proposition}[The generator realises the path]
\label{prop:path}
For every endpoint $y$, the rates of \eqref{eq:endpoint_gen}
satisfy the Kolmogorov forward equation of \eqref{eq:path}, and
their positionwise product realises the corresponding product path
over sequences.
\end{proposition}
Proposition~\ref{prop:path} is what makes the training corruption and
the sampling dynamics the same process: the prior is trained by
corrupting with $q_t(\cdot\mid z^\star)$ and predicting $z^\star$, and
the sampler then integrates the generator that path induces.

A masked transformer predicts, at each grid time, a clean-token
posterior per position, with classifier-free
guidance~\citep{ho2022cfg} and top-$k$ truncation; averaging
\eqref{eq:endpoint_gen} under this posterior gives the controlled
rate, which we realise with one hard endpoint sample per position.
Freezing the sampled rates over the interval $h_n$ and taking the
first event gives the parallel kernel
\begin{equation}
  P^{\mathrm{next}}_{n,\ell}(k)=
  \begin{cases}
    \exp[-h_nH_{n,\ell}], & k=i,\\[1mm]
    \bigl(1-\exp[-h_nH_{n,\ell}]\bigr)\,u_{n,\ell}(k)/H_{n,\ell}, & k\neq i,
  \end{cases}
  \label{eq:first_event}
\end{equation}
with $u_{n,\ell}(k)$ the sampled rates out of the current token $i$,
$H_{n,\ell}$ their sum, and all mass on $i$ when $H_{n,\ell}=0$ ---
the absorbed case.

\begin{proposition}[Tau-leap consistency]
\label{prop:tauleap}
The product of \eqref{eq:first_event} over positions agrees with
the posterior-averaged joint generator to first order in $h_n$, with
remainder bounded by $Ch_n^{2}\,\E[\mathcal H_n^{2}]$ for
$\mathcal H_n=\sum_\ell H_{n,\ell}$.
\end{proposition}
The generator the bound refers to is the controlled one, whose clean
posterior guidance and truncation tilt. Its remainder grows with the
sequence length and with $\dot\beta$, which rises towards $t{=}1$, so
the kernel behaves as a small-step integrator through the interior of
the grid and as a commitment to the sampled endpoint at the last
steps. Proofs are in Appendix~\ref{app:ctmc}.

\noindent\textbf{Anchor conditioning.} The model sees anchors as a
state-dependent residual: at each step the current clean posterior is
soft-decoded to joint positions $\hat{\vp}$, and the masked residual
$\vR_{\mathcal A}=\vM\odot(\vY-\hat{\vp})$ --- targets $\vY$, binary
observation mask $\vM$ --- is concatenated per frame with $\vY$ and $\vM$,
encoded by a temporal convolutional network, and injected through zero-initialised
low-rank adapters as additional keys and values in every attention block. The
conditioning is a feedback signal on the error of the model's own current
clean prediction.

\noindent\textbf{Solver-coupled training.} The VQ encoder, decoder, codebook
and text encoder stay frozen throughout. Training runs in three
stages: a text-conditioned prior trained with clean-token
cross-entropy over the path of \eqref{eq:path}; control fine-tuning under the
conditioning above, over a ladder of anchor densities with first and last
frames always observed; and a short on-policy calibration in which the model
generates with its own sampler while the terminal solve sits inside the
training graph. There the sampled prefix carries no graph, the last few states
stay differentiable through hard-forward
transitions~\citep{jang2017gumbel,maddison2017concrete}, and the solved state
re-enters the loss through the identity surrogate
\begin{equation}
  \vE^{\mathrm{ST}}
  =\vE^{\mathrm{diff}}
  +\operatorname{sg}\!\left[\vE^{\mathrm{num}}-\vE^{\mathrm{diff}}\right],
  \label{eq:st}
\end{equation}
whose forward value is the exact numerical solve and whose Jacobian is
the identity. Differentiating through \eqref{eq:gn_update} at frozen $\vJ$ gives
$\vI-\vJ^{\top}(\vJ\vJ^{\top}+\mu\vI)^{-1}\vJ$, which tends to the
orthogonal projector onto $\ker\vJ$, so the exact path would annihilate
precisely the gradient component along the correction just applied ---
the component that reports how far the emitted state was from meeting
the anchors.

The anchors act twice and the solver once: they condition every sampler
step, and at the last one the terminal clean state is handed to \eqref{eq:gn_update}
for $R{=}2$ accepted steps under the identity metric, after which the corrected
continuous state is decoded as it stands and never re-quantised
(Algorithm~\ref{alg:inloop}). What the state it emits then costs to correct is
\S\ref{sec:ours}.

\section{Experiments}
\label{sec:exp}

\begin{table}[t]
\centering
\caption{Control accuracy on the seven retrofitted hosts and our own
generator (HumanML3D validation,
$n{=}1024$, $K{=}8$, six joints), each against its own released
control method on identical clips, anchors and numerics. Depth $0$ is
the state each sweep starts from; solve depth
is the budget dial, reported whole. MoMask and MARDM release no
sparse-control method. Our generator carries the solver in its
sampler, so its incumbent column is the deployed output and the sweep
a second solver on top.}
\label{tab:crossmodel}
\footnotesize
\setlength{\tabcolsep}{2pt}
\begin{tabular}{@{}llc cccccc@{}}
\toprule
&&& \multicolumn{6}{c}{\methodname{} ctrl.\ err (m) at depth} \\
\cmidrule(lr){4-9}
Host & Its control method & Its err (m) & 0 & 1 & 2 & 3 & 5 & 8 \\
\midrule
MDM~\citep{tevet2023mdm} & DNO-800 & 0.0379 & 0.8920 & 0.1577 & 0.0417 & 0.0096 & 0.0001 & 0.0000 \\
MaskControl & branch + token opt. & 0.0053 & 0.4281 & 0.1533 & 0.0550 & 0.0193 & 0.0026 & 0.0002 \\
OmniControl & guidance & 0.1457 & 0.7445 & 0.1568 & 0.0474 & 0.0128 & 0.0006 & 0.0000 \\
CondMDI & keyframe cond. & 0.1260 & 0.7723 & 0.0772 & 0.0237 & 0.0074 & 0.0006 & 0.0001 \\
MotionLCM & CLT-400 & 0.0694 & 0.8120 & 0.4293 & 0.2707 & 0.1803 & 0.1006 & 0.0664 \\
\midrule
MoMask & --- & --- & 0.7649 & 0.3470 & 0.1437 & 0.0556 & 0.0084 & 0.0011 \\
MARDM~\citep{meng2025mardm} & --- & --- & 0.8499 & 0.2069 & 0.0545 & 0.0117 & 0.0003 & 0.0000 \\
\midrule
Ours & in-sampler solve & 0.0827 & 0.0827 & 0.0374 & 0.0180 & 0.0091 & 0.0026 & 0.0006 \\
\bottomrule
\end{tabular}
\end{table}

\begin{table}[t]
\centering
\caption{What the correction costs: Ctrl, Jerk and both skate ratios read at the shallowest recorded depth at which that host reaches $1$\,cm, HF and Displ.\ at depth three (Appendix~\ref{app:repro}).
$w_{\text{eff}}$ is the measured temporal support, the
number of frames moved by a change of one state coordinate, a property
of the frozen map $\mathcal D$ alone. Skate is given under both measured
conventions: mean foot speed during
contact~\citep{rempe2026kimodo}, and the fraction of contact frames
past the threshold~\citep{xie2024omnicontrol}. MaskControl decodes through the
frozen MoMask codebook and shares its $w_{\text{eff}}$; its HF and Displ.\ were not measured;
Ctrl and both skate ratios are at $n{=}1024$, jerk, HF and Displ.\ at $n{=}32$.}
\label{tab:families}
\footnotesize
\begin{tabular}{@{}ll cc c ccc cc@{}}
\toprule
& & & & & & & & \multicolumn{2}{c}{Skate ratio} \\
\cmidrule(lr){9-10}
Host & State & $w_{\text{eff}}$ & Depth & Ctrl (m) & Jerk & HF & Displ.\ (m) & speed & count \\
\midrule
CondMDI & abs.\ root & 1.0 & 3 & 0.0074 & $51.9\times$ & 0.487 & 0.039 & $6.21\times$ & $1.84\times$ \\
OmniControl & rel.\ root & 1.0 & 5 & 0.0006 & $27.6\times$ & 0.049 & 0.575 & $5.34\times$ & $4.76\times$ \\
MDM & rel.\ root & 1.0 & 3 & 0.0096 & $21.3\times$ & 0.077 & 0.672 & $4.84\times$ & $5.25\times$ \\
\midrule
MARDM & AE latent & 28.7 & 5 & 0.0003 & $4.6\times$ & 0.008 & 0.747 & $2.74\times$ & $4.67\times$ \\
MoMask & VQ token & 49.6 & 5 & 0.0084 & $2.0\times$ & 0.011 & 0.458 & $1.64\times$ & $2.06\times$ \\
MaskControl & VQ token & 49.6 & 5 & 0.0026 & $1.6\times$ & --- & --- & $1.26\times$ & $1.39\times$ \\
MotionLCM & VAE latent & 54.2 & 40 & 0.0073 & $2.4\times$ & 0.008 & 0.863 & $1.30\times$ & $1.37\times$ \\
\midrule
Ours & VQ token & 43.1 & 3 & 0.0091 & $1.1\times$ & 0.008 & 0.098 & $1.12\times$ & $1.29\times$ \\
\bottomrule
\end{tabular}
\end{table}

\begin{table}[!t]
\centering
\caption{HumanML3D test split, six-point control. The upper block is uncontrolled
generation, where every system performs the same text-to-motion task.
The lower block is the same six-point anchor protocol throughout, so
generation quality is read under a fixed control task. Control
accuracy is Table~\ref{tab:crossmodel} (validation split). Foot skate is the fraction of contact frames past the threshold~\citep{xie2024omnicontrol}. $\rightarrow$: closer to real is better.}
\label{tab:published}
\footnotesize
\begin{tabular}{@{}lcccc@{}}
\toprule
Method & FID $\downarrow$ & Top-3 $\uparrow$ & Diversity $\rightarrow$ & Foot skate $\downarrow$ \\
\midrule
\multicolumn{5}{@{}l}{\emph{Uncontrolled generation}} \\
Real & 0.002 & 0.797 & 9.503 & --- \\
MoMask~\citep{guo2024momask} & 0.082 & 0.797 & --- & --- \\
MOGO~\citep{fu2026mogo} & 0.064 & 0.801 & --- & --- \\
Motion Anything~\citep{zhang2025motionanything} & 0.028 & 0.829 & --- & --- \\
MoGeFlow~\citep{fang2026mogeflow} & 0.058 & 0.873 & 9.981 & --- \\
CTMC prior (ours) & 0.043 & 0.827 & 9.669 & --- \\
\midrule
\multicolumn{5}{@{}l}{\emph{Six-point sparse control}} \\
CondMDI~\citep{cohan2024condmdi} & 0.498 & 0.631 & 9.013 & 0.099 \\
OmniControl~\citep{xie2024omnicontrol} & 0.689 & 0.689 & 9.381 & 0.091 \\
TLControl~\citep{wan2024tlcontrol} & 4.637 & 0.473 & 8.078 & 0.058 \\
MotionLCM~\citep{dai2024motionlcm} & 1.013 & 0.676 & 8.722 & 0.141 \\
MaskControl~\citep{pinyoanuntapong2025maskcontrol} & 0.116 & 0.842 & 8.940 & 0.082 \\
\midrule
\methodname{} (ours) at $0.083$\,m & 0.102 & 0.796 & 9.598 & 0.057 \\
\methodname{} (ours) at $0.0026$\,m & 0.102 & 0.799 & 9.038 & 0.083 \\
\bottomrule
\end{tabular}
\end{table}

\subsection{Control accuracy across hosts}
\label{sec:crossmodel}

Seven published generators expose the interface of \S\ref{sec:problem}: MDM~\citep{tevet2023mdm},
OmniControl~\citep{xie2024omnicontrol} and
CondMDI~\citep{cohan2024condmdi} on motion features,
MoMask~\citep{guo2024momask} and MaskControl~\citep{pinyoanuntapong2025maskcontrol} on VQ token embeddings,
MARDM~\citep{meng2025mardm} on a continuous autoencoder latent, and
MotionLCM~\citep{dai2024motionlcm} on a sequence-VAE latent
(Table~\ref{tab:families}). Every sweep starts from the host's text-conditioned sample with its control machinery off, depth $0$ in Table~\ref{tab:crossmodel} (MaskControl from its control branch, Appendix~\ref{app:repro}). All experiments use
HumanML3D~\citep{guo2022humanml3d} with six control joints (root,
head, both wrists, both feet) at $K$ anchor frames placed uniformly
including the endpoints; $K{=}8$ unless stated, giving $A=48$ anchors and $m=144$ residual coordinates; control error is flat out to $K{=}32$, $m{=}576$ (Appendix~\ref{app:density}). Targets come from the held-out clip; control error is the mean
anchor-to-decoded-joint distance. A conformance gate checks clips, anchors,
solver parameters and numerics across hosts (Appendix~\ref{app:repro}).

In anchor error the sweep dominates or matches the incumbent on every host that
releases a control method, and closes the anchors on the two that do
not (Table~\ref{tab:crossmodel}). MotionLCM needs the most depth, the one host where the disproportion of \S\ref{sec:problem} thins (Appendix~\ref{app:motionlcm}). Wall-clock at the depth that matches the incumbent, generation included: MDM $2.6$\,s against DNO's $172.3$\,s, MaskControl $2.0$ against $9.9$, OmniControl $40.9$ against $81.4$, CondMDI $36.9$ against $35.4$, MotionLCM $11.8$ against CLT-400's $13.1$; MoMask $1.9$ and MARDM $14.0$ to a centimetre. At matched anchor error the quality cost follows the decoder family as the physical cost does; our generator closes its anchors at FID $0.078$ in $0.50$\,s, below MaskControl's and MotionLCM's released methods (Appendix~\ref{app:repro}, Table~\ref{tab:quality_incumbent}).

The last row is our own generator under the same operator: it emits
$0.083$\,m and three accepted steps take it to $0.0091$\,m. One damping constant covers all eight,
which rests on the scale normalisation of \S\ref{sec:damping}
(Appendix~\ref{app:solver}).


\subsection{Where the physical cost comes from}
\label{sec:law}

Jerk at the anchored frames measures the mechanism directly, and it
separates the windowed decodes from the analytic recoveries without
overlap (Table~\ref{tab:families}): $21$--$52\times$ against $1.1$--$4.6\times$, and MARDM at $0.0003$\,m pays a fifth of what MDM pays at $0.0096$\,m. The measured temporal support divides them the same
way, an order of magnitude apart: $1.0$ frame on the analytic recoveries, $28.7$ to $54.2$ on the learned decoders.

CondMDI sits where \S\ref{sec:coupling} places the sharp limit:
absolute-root and frame-local, its displacement confined to
the anchored frames in every clip, and it pays the most. MDM and OmniControl integrate the root, so on their root channels the correction is a prefix velocity bias --- a rigid drift, not a deformation --- which moves the body $0.672$ and $0.575$\,m, and they still pay $21$--$28\times$. Down-weighting those channels in the solver's metric leaves the cost where it is (Appendix~\ref{app:metric}).

The size of the correction runs the other way from its cost:
MotionLCM moves the motion $0.863$\,m and MARDM $0.747$\,m, the two
largest displacements measured, at jerk ratios of $2.4$ and $4.6$,
while CondMDI moves it $0.039$\,m at $51.9$, the highest.

The downstream cost in foot skating depends on which convention
measures it, and the dependence is the mechanism again. Under mean
foot speed during contact the families separate as jerk does,
$4.84$--$6.21\times$ against $1.12$--$2.74\times$; under the fraction
of contact frames above the threshold they do not, CondMDI falling to
$1.84\times$ and MARDM rising to $4.67\times$: a correction confined to the
anchored frames makes a few very fast slides, which a mean registers and a
count barely does, and one across the clip makes many slow ones.

At matched accuracy our generator closes its anchors to $0.0091$\,m
and pays $1.12\times$ at $w_{\text{eff}}{=}43.1$ against its deployed output and $1.37\times$ against its raw sampler output; the separation holds on absolutes as on ratios (Appendix~\ref{app:repro}).

Reparameterising the update through a temporal basis of width $w$
tests the same variable causally, at fixed weights, decoder, anchors
and solver (Table~\ref{tab:basis}, Appendix~\ref{app:basis}). Where the decoder is
frame-local the basis supplies that support from outside and the
physical cost falls monotonically with it, skating $3.87$ to
$3.15\times$ on MDM and $5.40$ to $4.30\times$ on OmniControl from $w{=}1$ to $8$, read at the first depth under a centimetre ($n{=}128$); at $w{=}16$ neither host gets there. On MoMask,
whose decoder already carries $w_{\text{eff}}{=}49.6$, the basis buys
nothing: the high-frequency share stays at $0.009$--$0.011$, skating
at a centimetre at $1.6$--$1.8\times$ over $w{=}1$ to $4$, and accuracy
falls, $2.1\times$ and $3.4\times$ at depth three for $w{=}8$ and $16$,
neither reaching a centimetre. An explicit basis supplies a smoothness
the windowed decoder already has, and pays for it in the directions
the solve had.


\subsection{Our generator}
\label{sec:ours}

\begin{figure}[t]
\centering
\includegraphics[width=0.7\textwidth]{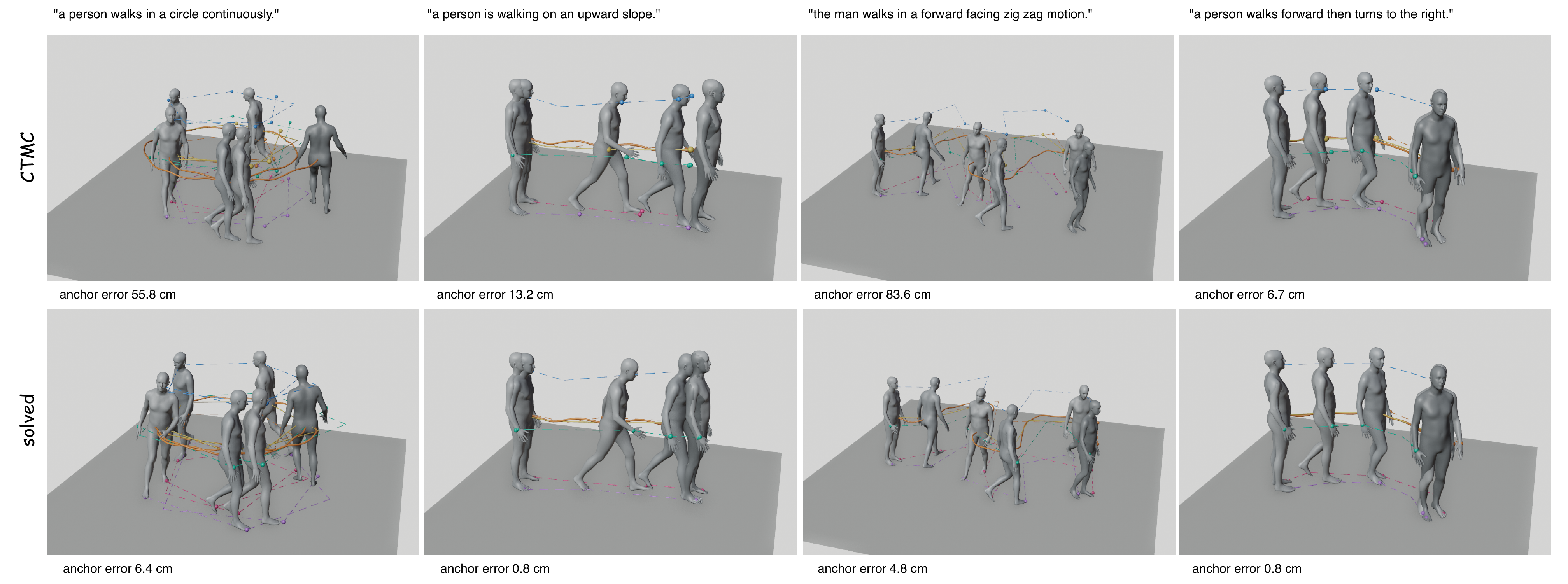}
\caption{Four requests under the six-point protocol, $K{=}8$. Spheres
are the anchor targets, one colour per joint, drawn at the frames they
constrain; the bodies are the generated motion at those same frames.
Above, the sample the sampler emits; below, the same sample after the
terminal solve, which carries the body onto the anchors from as far as
$0.84$\,m.}
\label{fig:qual}
\end{figure}

\noindent\textbf{Deployed results.} The deployed configuration is
Algorithm~\ref{alg:inloop}: $N{=}20$ sampler steps, one terminal
solve, $R{=}2$ accepted steps. On the HumanML3D test split at $K{=}8$
it reaches $0.083$\,m control error at FID $0.102$, Top-3 $0.796$ and
diversity $9.60$; latency is $0.50$\,s per clip unbatched
(validation, $n{=}64$); three further accepted steps take it to $0.0091$\,m in $1.5$\,s and five to $0.0026$\,m in $2.1$\,s.
Under the six-point protocol (Table~\ref{tab:published}) that is the
best FID in the block, ahead of MaskControl's $0.116$, with the
diversity closest to real motion of any row and foot skating $0.057$
against $0.082$. Solved to MaskControl's own accuracy, $0.0026$ against $0.0053$\,m, FID holds at $0.102$ against $0.116$, skating meets theirs at $0.083$ against $0.082$, and Top-3 is $0.799$ against $0.842$. The upper block places the generator
before the solve: the prior scores FID $0.043$ at Top-3 $0.827$.

Figure~\ref{fig:qual} shows what those two steps do: the sampler
places the body $0.067$--$0.836$\,m from the anchors and the solve
brings it to $0.008$--$0.064$\,m, closing the largest gap by
$17\times$.

Trained with $R{=}1$ instead of $R{=}2$, the same recipe deploys to
$0.166$ and $0.087$\,m at the two inference depths against the deployed
checkpoint's $0.165$ and $0.083$ (validation, $n{=}1024$): the second step
buys what a step of the solver buys. Trained post hoc instead, with no solver in its sampler, the same model emits $0.367$\,m against $0.314$ and stands at $0.107$\,m after the same two steps (Appendix~\ref{app:training}).

\section{Conclusion}
\label{sec:conclusion}

Sparse anchors fix a few hundred directions of a state with tens of thousands;
what they leave open is the rest. A training-free damped Gauss--Newton solve of
the anchor residual, one dimensionless damping constant and no per-host tuning,
takes the increment that moves none of it, and in anchor error it matches or
exceeds every released control method it is measured against.

What a correction may then look like is the decoder's to decide, and holding
the solver fixed measures what its shapes cost: at matched anchor accuracy the
decoder families separate without overlap in jerk and in mean contact speed.
Built to that criterion, our generator reaches $0.083$\,m at FID $0.102$.
Temporal support is measurable in advance on any host that exposes a differentiable decode, and is a design choice.

\section*{Ethics statement}

This work synthesises human motion from text and Cartesian anchors on
public HumanML3D data, and releases no new human data. The generators
it retrofits are public released checkpoints, used under their own
licences.

\section*{Reproducibility statement}

The solver is \eqref{eq:gn_update} with the schedule and constants
of Appendix~\ref{app:solver}; the protocol, conformance gate and
metric conventions are in Appendix~\ref{app:repro}; the generator's
three training stages are in Appendix~\ref{app:training}. Code, the
gate's output and the per-sample records will accompany the paper.

\appendix

\section{Solver details}
\label{app:solver}

\noindent\textbf{The loop, in full.} At outer step $k$ the solver holds a
state $\vE_k$, evaluates $\vr_k=\vr(\vE_k)$ and builds
$\vJ_k$ by $m$ reverse-mode products through $\mathcal D$. A trial
uses $\mu_k=\lambda_k\,\tfrac1m\operatorname{tr}(\vJ_k\vJ_k^{\top})$ in
\eqref{eq:gn_update}, and is accepted when
$\Phi(\vE_k+\bDelta)<\Phi(\vE_k)$ strictly, for the merit $\Phi$ of \S\ref{sec:gn}. On rejection
$\lambda\leftarrow10\lambda$ and the same linear system is re-solved
with the new damping; the Jacobian is rebuilt only after an accepted
step. Up to $16$ trials are allowed per step on the retrofitted hosts, and
four in the deployed configuration of \S\ref{sec:ours}. On acceptance
$\lambda\leftarrow\max(\lambda/2,10^{-6})$ and is carried into the next
step rather than reset. The solve terminates when the mean squared residual
$\Phi/m$ falls below $10^{-10}$, the deployed configuration of
\S\ref{sec:ours} using $10^{-5}$ in its place (Appendix~\ref{app:training});
when a step exhausts its trials, when
$\operatorname{tr}(\vJ\vJ^{\top})$ underflows, or at the step budget;
it returns the last accepted state. The $m\times m$ system is solved
by LU with partial pivoting in single precision with TF32 disabled;
a non-finite solution is treated as a rejected trial.

\noindent\textbf{What one accepted step costs.} Building $\vJ$ costs $m$
reverse-mode products through the frozen decoder; forming
$\vJ\vJ^{\top}$ costs $O(m^2P)$; the solve costs $O(m^3)$; recovering
the primal step costs $O(mP)$. With $m=144$ and
$P\in[10^4,10^5]$ the decoder passes dominate. A rejected trial adds
one $O(m^3)$ re-solve and nothing else.

\noindent\textbf{Reported quantities.} Control error is the mean over
anchors of the Euclidean distance between the decoded joint and its
target, in metres; it is a different functional from the merit
$\Phi$. Relative state motion is
$\lVert\vE_R-\vE_0\rVert_2/\lVert\vE_0\rVert_2$, a latent-space quantity. Steps-to-a-centimetre is the first accepted step at
which control error falls below $10^{-2}$\,m.

\noindent\textbf{Scale normalisation.} Replacing
\eqref{eq:scale_damping} by the same \emph{absolute} damping on
every host breaks the transfer: step-3 control error ($n{=}32$) rises
from $0.0435$ to $0.3920$\,m on MoMask and from $0.0017$ to
$0.0166$\,m on MDM, while MotionLCM, whose Gram scale sits near unity,
is unchanged at $0.204$\,m.

\noindent\textbf{Gauss--Newton against first-order descent.} The
same state, decoder, residual and clips admit any optimiser, so we
ran Adam on $\tfrac12\lVert\vr\rVert^2$ through the frozen decoder
alone, on MDM and MoMask (validation, $n{=}64$), with the learning
rate set to $\eta$ times the standard deviation of the state,
$\eta\in\{0.01,0.03,0.1\}$, $400$ iterations, and the best $\eta$
chosen after the fact; timing covers the update only, as for the
solver, in a run separate from the timings of \S\ref{sec:crossmodel}. On MDM, Adam stands at $0.041$\,m after $0.46$\,s and
$0.016$\,m after $0.92$\,s, where the solver stands at $0.038$\,m
(two steps, $0.66$\,s) and $0.010$\,m (three, $0.99$\,s); at
$1.3$\,s the solver is below $10^{-4}$\,m at depth five and Adam is
at $0.006$\,m after $1.85$\,s. On MoMask, Adam is ahead early,
$0.024$\,m at $0.79$\,s against the solver's $0.131$\,m at two
steps, the two meet at a centimetre ($0.011$\,m at $1.56$\,s against
$0.011$\,m at depth five, $1.95$\,s), and below it the solver pulls
away, $0.0002$\,m at depth sixteen ($3.5$\,s) against $0.005$\,m
after $400$ iterations ($3.1$\,s). Mean contact-speed skating is the
same for both optimisers at matched error, $4.5$--$4.8\times$ on MDM
and $1.5$--$1.6\times$ on MoMask: the physical cost belongs to the
decoder, not to the optimiser. What the curvature buys is the last
order of magnitude, from the centimetre down, at equal wall-clock.

\section{Choice of metric}
\label{app:metric}

Replacing the identity by $\mSigma$ in \eqref{eq:gn_objective} gives
$\bDelta=-\mSigma\vJ^{\top}(\vJ\mSigma\vJ^{\top}+\mu\vI)^{-1}\vr$. We
compared the identity against a diagonal state covariance and the full
codebook covariance on our own generator, sweeping solve depth on the
validation split (Table~\ref{tab:metric}). Identity and diagonal track
each other at every depth on both axes. The codebook covariance reaches
the anchors more slowly and costs a factor of four in FID from the
first step onward, so we use the identity throughout.

Depth zero is the untouched generation, and its FID of $0.171$ is the
highest in the identity column: the first accepted step reduces FID to $0.090$
while cutting control error by $46\%$, and FID then stays below the
untouched value.

\begin{table}[h]
\centering
\caption{Solve depth against solver metric, on the post-hoc-trained
checkpoint whose sampler carries no solver, validation split, $n{=}1504$. Identity
and diagonal track each other on both axes; the codebook covariance
costs a factor of four in FID from the first step.}
\label{tab:metric}
\small
\begin{tabular}{@{}l cc cc cc@{}}
\toprule
& \multicolumn{2}{c}{Identity} & \multicolumn{2}{c}{Diagonal} & \multicolumn{2}{c}{Codebook cov.} \\
\cmidrule(lr){2-3} \cmidrule(lr){4-5} \cmidrule(lr){6-7}
Depth & Ctrl (m) & FID & Ctrl (m) & FID & Ctrl (m) & FID \\
\midrule
0 & 0.3573 & 0.1714 & 0.3573 & 0.1714 & 0.3573 & 0.1714 \\
1 & 0.1943 & \textbf{0.0901} & 0.1906 & \textbf{0.0888} & 0.2047 & 0.3571 \\
2 & 0.1031 & 0.0973 & 0.0994 & 0.0957 & 0.1004 & 0.3499 \\
3 & 0.0523 & 0.0984 & 0.0486 & 0.0959 & 0.0572 & 0.3776 \\
4 & 0.0257 & 0.1003 & 0.0240 & 0.0983 & 0.0377 & 0.3933 \\
5 & \textbf{0.0129} & 0.1040 & \textbf{0.0119} & 0.1026 & 0.0259 & 0.4111 \\
\bottomrule
\end{tabular}
\end{table}

The metric also fixes which coordinates a correction is cheap in, so it
is the second place the shape of \S\ref{sec:coupling} could come from.
On the two hosts whose increment spreads over the prefix we varied it
directly: $\mSigma=\operatorname{diag}(\vw)$ with $\vw$ reduced on the
integrated root channels and unity elsewhere, every row read at the
shallowest recorded depth reaching $1$\,cm, validation split, $n{=}32$.
Reducing those channels tenfold moves the contact-speed ratio by under
one percent --- $3.89$ to $3.87\times$ on MDM, $5.43$ to $5.44\times$
on OmniControl --- and jerk by under two percent, while the solve takes
two further accepted steps to reach the same centimetre. The cost
stays where the decoder puts it.

\section{Anchor density}
\label{app:density}

\begin{table}[h]
\centering
\caption{Anchor density on the test split. $K{=}8$ sits
exactly at the shipped Jacobian row cap of $48$ pairs; $K{=}16$ and
$K{=}32$ use an eval-time cap of $192$ so that every cell has full
coverage. Match is the Guo matching score, the mean text--motion
embedding distance, for which real motion scores $2.974$.}
\label{tab:density}
\small
\begin{tabular}{@{}lccccccc@{}}
\toprule
$K$ & Ctrl (m) & FID & Top-1 & Top-3 & Diversity & Foot skate & Match \\
\midrule
2 & 0.0461 & 0.2294 & 0.4907 & 0.7720 & 9.8485 & 0.0438 & 3.1195 \\
4 & 0.0832 & 0.1511 & 0.4955 & 0.7843 & 9.7391 & 0.0523 & 3.0573 \\
8 & 0.0830 & 0.1016 & 0.5119 & 0.7955 & 9.5975 & 0.0570 & 3.0209 \\
16 & 0.0845 & 0.0674 & 0.5112 & 0.7881 & 9.6312 & 0.0588 & 3.0054 \\
32 & 0.0797 & 0.0580 & 0.5112 & 0.7931 & 9.5536 & 0.0558 & 3.0159 \\
\bottomrule
\end{tabular}
\end{table}

Control error is flat from $K{=}4$ to $K{=}32$, inside $0.079$--$0.085$\,m, while the residual dimension $m=18K$ grows from $72$ to $576$: the solve holds as the constraint set grows by a factor of eight. Across the whole sweep, from $K{=}2$, Top-3 spans $0.772$--$0.796$ and diversity $9.55$--$9.85$.

\section{Temporal basis of the update}
\label{app:basis}

The update is confined to a temporal basis of width $w$ frames --- the
increment for each state channel is the linear interpolation of
$\lceil T/w\rceil$ knots placed $w$ frames apart, the first and last
on the first and last frame, so each knot carries a triangular bump of
support $2w$ frames; the basis has rank $C\lceil T/w\rceil$ for $C$
channels, carries no normalisation, and at $w{=}1$ is the identity, so
that arm is the unmodified solve. On the frame-local hosts $w$ is in
motion frames; on MoMask the basis lives on the token axis of its
stride-four tokeniser, so $w$ is in tokens, $4w$ frames. Everything else is held: weights, frozen decoder,
clips, anchors, damping and accept schedule. Reading the intervention
against the temporal support the host's own decoder already carries
gives the two directions in \S\ref{sec:law}.

\begin{table}[h]
\centering
\caption{Intervention on a fixed host ($n{=}128$): the update
reparameterised through a temporal basis of width $w$ (frames on MDM and OmniControl, stride-four tokens on MoMask); weights,
decoder, anchors and solver unchanged. Each host is labelled with the
temporal support its own decoder already carries. Ctrl is a ratio to that host's own
$w{=}1$ arm, in which the basis is the identity and the solve is
the unmodified one, read mid-descent at depth three; Final err is the error at the end of the sixteen-step sweep; Skate is the ratio of mean contact speed to the unconstrained generation, read at the first recorded depth at which that width's error is under a centimetre (--- where the sweep never reaches it); HF is absolute, read at the end of the sweep.}
\label{tab:basis}
\small
\begin{tabular}{@{}ll ccccc@{}}
\toprule
& & \multicolumn{5}{c}{basis width $w$} \\
\cmidrule(lr){3-7}
Host & & 1 & 2 & 4 & 8 & 16 \\
\midrule
MDM & Ctrl & $1.00\times$ & $0.73\times$ & $0.69\times$ & $0.89\times$ & $3.09\times$ \\
$w_{\text{eff}}{=}1.0$ & Final err (m) & 0.0000 & 0.0000 & 0.0000 & 0.0014 & 0.0251 \\
 & Skate at $1$\,cm & $3.87\times$ & $3.71\times$ & $3.48\times$ & $3.15\times$ & --- \\
 & HF & 0.062 & 0.016 & 0.012 & 0.012 & 0.012 \\
\midrule
OmniControl & Ctrl & $1.00\times$ & $0.31\times$ & $0.33\times$ & $0.66\times$ & $2.49\times$ \\
$w_{\text{eff}}{=}1.0$ & Final err (m) & 0.0000 & 0.0000 & 0.0000 & 0.0003 & 0.0138 \\
 & Skate at $1$\,cm & $5.40\times$ & $5.22\times$ & $4.88\times$ & $4.30\times$ & --- \\
 & HF & 0.045 & 0.013 & 0.010 & 0.010 & 0.011 \\
\midrule
MoMask & Ctrl & $1.00\times$ & $0.94\times$ & $1.23\times$ & $2.13\times$ & $3.36\times$ \\
$w_{\text{eff}}{=}49.6$ & Final err (m) & 0.0001 & 0.0001 & 0.0013 & 0.0111 & 0.0472 \\
 & Skate at $1$\,cm & $1.67\times$ & $1.58\times$ & $1.79\times$ & --- & --- \\
 & HF & 0.009 & 0.009 & 0.010 & 0.010 & 0.011 \\
\bottomrule
\end{tabular}
\end{table}

\section{The one host where the disproportion thins}
\label{app:motionlcm}

\S\ref{sec:problem} rests on a disproportion: a few dozen scalar
constraints against a state of $10^4$--$10^5$ dimensions, which is
what makes the projection small. MotionLCM is the one host in this
study where that margin closes. Its sequence-VAE latent is $512$
dimensions, and the six-point request at $K{=}8$ fixes $144$ of them
--- $28$ percent, against a state of tens of thousands of coordinates on every other host.

The solve behaves as the disproportion predicts. It reaches $1$\,cm at
depth $40$, where CondMDI, MDM and our own generator reach it at
depth three and OmniControl, MARDM, MoMask and MaskControl at depth five; the damping schedule accepts at every depth and the residual falls monotonically, $0.812$\,m at depth zero to $0.0664$ at $8$, $0.0113$ at $32$, $0.0073$ at $40$ and $0.0053$ at $48$. What it
buys per step is simply smaller, and each Jacobian row costs a VAE
decode, so depth $40$ takes $30.0$\,s of solve against a generation
that takes $0.015$\,s.

The
host's released control method stops at $0.0694$\,m (Table~\ref{tab:crossmodel}), its $400$ latent-optimisation steps taking $13.1$\,s on the timing protocol of \S\ref{sec:crossmodel};
the solve reaches $0.0073$\,m, nine times nearer, in $30.0$\,s.

\section{Reproducibility}
\label{app:repro}

\noindent\textbf{How Table~\ref{tab:families} is read.} Jerk is the
ratio of jerk at the anchored frames to the unconstrained generation.
HF share is the fraction of the per-frame displacement magnitude's
power spectrum lying above $5$\,Hz, and Displ.\ is that magnitude's mean; both are absolute. Jerk and the contact-speed skate ratio are taken against that host's own unconstrained generation, whose absolute values are, in m/s of contact speed and in metres per frame cubed at $20$\,fps: CondMDI $0.251$ and $0.0169$, OmniControl $0.135$ and $0.0035$,
MDM $0.163$ and $0.0045$, MARDM $0.249$ and $0.0077$, MoMask $0.196$
and $0.0077$, MaskControl $0.182$ and $0.0070$, MotionLCM $0.271$ and
$0.0082$, ours $0.201$ and $0.0102$. Ours is the deployed output, already corrected by its own two accepted steps; against the state the sampler emits before that solve (Appendix~\ref{app:training}) the contact-speed ratio at depth three is $1.37\times$ and the count ratio $2.03\times$. Read as absolutes at the matched
depth the two groups still separate without overlap, at
$0.72$--$1.56$ against $0.23$--$0.68$ in skate and $0.097$--$0.879$
against $0.011$--$0.035$ in jerk, with ours among the lowest on both.
Ctrl and both skate ratios are at $n{=}1024$; jerk, HF and Displ.\ at $n{=}32$;
$w_{\text{eff}}$ at $16$ clips with $32$ probes each; on a learned decoder every state coordinate feeds every feature channel of the frames in its window, the integrated root among them, so the reach reported there includes the recovery. Every row is
taken at the shallowest recorded depth reaching $1$\,cm, which is three for
CondMDI, MDM and our generator, five for OmniControl, MARDM, MoMask and MaskControl, and forty for MotionLCM. HF and Displ.\ move by under three percent between depth three and depth sixteen, and are read at depth three, at sixteen for MotionLCM, the deepest depth at which they were recorded.

\noindent\textbf{Quality against the released control method.} Table~\ref{tab:quality_incumbent} scores three hosts under one evaluator, every row, generated and reference alike, decoded to joints and its $263$-d features re-extracted from them, so the evaluator sees only what the control moved: the host's unconstrained sample, its released control method, and \methodname{} at the first recorded depth that matches the released method's anchor error, with the wall-clock of \S\ref{sec:crossmodel}. The quality cost orders by decoder family as the physical cost of Table~\ref{tab:families} does. On the windowed host \methodname{} reaches half the released method's anchor error in a fifth of its time at equal Top-3, paying FID $0.236$ against $0.103$; on the VAE host it matches the error at parity in time and pays more; on the frame-local analytic host the correction is the sharp limit of \S\ref{sec:coupling}, and the quality goes with it. Our generator, built to the criterion and trained through the solve, closes its anchors at FID $0.078$ in $0.50$\,s, below the released control methods of MaskControl and MotionLCM at a fraction of their time.

\begin{table}[h]
\centering
\caption{Generation quality against each host's released control method (HumanML3D validation, one evaluator, features re-extracted from decoded joints for every row including the reference pool). Depth is the first recorded depth of Table~\ref{tab:crossmodel} at or under the released method's error; time is generation plus the solve to that depth, as in \S\ref{sec:crossmodel}. $n{=}1024$; CondMDI $n{=}256$, whose three rows are comparable with each other only. Ctrl is measured within this run; Table~\ref{tab:crossmodel} carries the campaign's values. Contact speed in m/s.}
\label{tab:quality_incumbent}
\footnotesize
\setlength{\tabcolsep}{2.5pt}
\begin{tabular}{@{}ll cccccc@{}}
\toprule
Host & Arm & Ctrl (m) & Time (s) & FID $\downarrow$ & Top-3 $\uparrow$ & Div. & Contact \\
\midrule
MaskControl & control branch only & 0.428 & --- & 0.049 & 0.800 & 9.59 & 0.182 \\
(windowed) & released, branch + opt.\ $100/600$ & 0.0050 & 9.9 & 0.103 & 0.813 & 9.31 & 0.206 \\
 & \methodname{}, depth $5$ & 0.0026 & 2.0 & 0.236 & 0.806 & 9.18 & 0.230 \\
\midrule
MotionLCM & unconstrained & 0.826 & --- & 0.256 & 0.833 & 9.25 & 0.271 \\
(VAE) & released: CLT-400 & 0.069 & 13.1 & 0.186 & 0.802 & 9.13 & 0.245 \\
 & \methodname{}, depth $8$ & 0.068 & 11.8 & 1.05 & 0.736 & 8.45 & 0.294 \\
\midrule
CondMDI & unconstrained (uncond.\ ckpt) & 0.751 & --- & 0.318 & 0.773 & 9.21 & 0.239 \\
(frame-local) & released, keyframe ckpt & 0.131 & 35.4 & 0.173 & 0.742 & 8.70 & 0.218 \\
 & \methodname{}, depth $1$ & 0.075 & 36.9 & 14.2 & 0.559 & 5.13 & 1.488 \\
\midrule
Ours (windowed, & deployed, $R{=}2$ & 0.083 & 0.50 & 0.078 & 0.798 & 9.36 & 0.201 \\
trained through solve) & & & & & & & \\
\bottomrule
\end{tabular}
\end{table}

\noindent\textbf{One protocol for every system.} All hosts receive the same
clip list, the same anchor frames and the same targets. An automated
conformance gate verifies byte-level equality of the clip list and the
anchor tensor, together with solver parameters, damping constant, row
cap, and the TF32 and seeding state, before any host is measured; its
output ships with the code. Timing is per host on one card, and the released-budget DNO measurements were taken on an H100 and the
rest on one A100. Clips are generated one at a time ($n{=}64$), and the sweep is one solve of sixteen accepted steps read at the depths of Table~\ref{tab:crossmodel} (MotionLCM's continues to depth $48$, Appendix~\ref{app:motionlcm}); the times of \S\ref{sec:crossmodel} are generation plus the solve up to the depth that matches the incumbent (MDM three, MaskControl five, OmniControl two, CondMDI one, MotionLCM eight; MoMask and MARDM to a centimetre, five), read from the same runs. Generation runs each host's released sampler: MDM at $100$ DDIM steps; OmniControl and CondMDI with their $1000$-step samplers, OmniControl at its fixed $196$ frames; MoMask and MaskControl's branch at $10$ masked steps, MaskControl's own method adding $100$ token-optimisation iterations per step and $600$ at the end; MARDM at $18$; MotionLCM at one. The starting states are the hosts' text-conditioned samples with their control machinery off: OmniControl without its spatial guidance, CondMDI's unconditional checkpoint, MotionLCM before its latent optimisation; MaskControl's sweep starts from its trained control branch's output with the token optimisation switched off, since that branch is the model.

\noindent\textbf{Noise floor.} Twenty repetitions of the deployed configuration give $95$ percent intervals of $\pm0.0036$ in FID, $\pm7\times10^{-4}$\,m in control error, $\pm4\times10^{-4}$ in foot skating and below $10^{-4}$ in jerk. On the retrofitted hosts, one configuration run twice on one card ($n{=}8$, identical seed) gives a spread in control error of $1.8\times10^{-4}$\,m
averaged over the depth grid, $1.2\times10^{-3}$\,m at depth three
where the solve is mid-descent and the accept--retry schedule can take
a different branch, and $9\times10^{-5}$\,m from depth eight onwards,
where the residual is small and the step is essentially determined.

\noindent\textbf{Metric conventions.} Foot skating is computed on the clip
truncated to its true length. Every skating number in this paper is computed under the truncated
convention on one evaluation stack.

\noindent\textbf{What the backward pass carries.} The solve enters the graph
through the straight-through surrogate of \eqref{eq:st}. Gradient
flows through the last few sampler states and the differentiable
suffix. Endpoint samples and suffix transitions are hard on the
forward pass; the discrete row selection of the next generator and the
truncated support of the guided posterior are held fixed in the
backward pass; the sampled prefix carries no graph. 

\section{Proofs}
\label{app:ctmc}

\subsection{Proposition~\ref{prop:path}}

Fix an endpoint $y$, write $\delta_k=\delta(k,y)$ and
$\bar\delta_t=\sum_iq_t(i\mid y)\delta_i$. Differentiating
\eqref{eq:path},
\begin{equation}
    \frac{\mathrm d}{\mathrm dt}q_t(k\mid y)
    =
    \dot\beta(t)\,q_t(k\mid y)\,[\bar\delta_t-\delta_k].
    \label{eq:path_derivative}
\end{equation}
For the rates of \eqref{eq:endpoint_gen}, using
$[a]_{+}-[-a]_{+}=a$,
\begin{equation}
    \begin{aligned}
    \sum_i q_t(i\mid y)\,Q_t^{y}(i,k)
    &=
    \sum_{i\neq k} q_t(i\mid y)\,q_t(k\mid y)\,
    \dot\beta(t)[\delta_i-\delta_k]_{+}
    -
    q_t(k\mid y)\sum_{j\neq k}q_t(j\mid y)\,
    \dot\beta(t)[\delta_k-\delta_j]_{+}
    \\
    &=
    \dot\beta(t)\,q_t(k\mid y)
    \sum_iq_t(i\mid y)[\delta_i-\delta_k]
    =
    \dot\beta(t)\,q_t(k\mid y)[\bar\delta_t-\delta_k],
    \end{aligned}
\end{equation}
which is \eqref{eq:path_derivative}. For an endpoint sequence
$\vy=(y_1,\dots,y_L)$ the product path
$q_t(\vx\mid\vy)=\prod_\ell q_t(x_\ell\mid y_\ell)$ and the
single-coordinate joint generator
$\mathcal Q_t^{\vy}(\vx,\vx^{\ell\to k})=Q_t^{y_\ell}(x_\ell,k)$,
with zero rate between sequences differing in more than one
coordinate, satisfy the corresponding forward equation: the time
derivative of the product is a sum of $L$ terms, each differentiating
one factor, and the joint generator moves probability only along
single-coordinate replacements, so the single-coordinate identity
applies termwise. \qed

\subsection{Proposition~\ref{prop:tauleap}}

Condition on the endpoint samples, so all rates are fixed. For one
position with current token $i$ and exit rate $H_{n,\ell}$, the
bounds $|e^{-x}-1+x|\le x^2/2$ and $|1-e^{-x}-x|\le x^2/2$ for
$x\ge0$ give
\begin{equation}
    P_{n,\ell}^{\mathrm{next}}
    =
    \delta_{i}
    +
    h_n\,Q_{\ell}(i,\cdot)
    +
    r_\ell,
    \qquad
    \lVert r_\ell\rVert_{\mathrm{TV}}
    \le
    (h_nH_{n,\ell})^2,
\end{equation}
where $Q_\ell(i,\cdot)$ places $u_{n,\ell}(k)$ on each $k\neq i$ and
$-H_{n,\ell}$ on $i$, so
$\lVert h_nQ_\ell(i,\cdot)\rVert_{\mathrm{TV}}=2h_nH_{n,\ell}$.
Expanding $\bigotimes_\ell P^{\mathrm{next}}_{n,\ell}$ around
$\bigotimes_\ell\delta_{x_\ell}$: the empty selection contributes
$f(\vx)$; the single selections contribute $h_n$ times the sampled
joint single-coordinate generator applied to $f$, plus $r_\ell$ terms
bounded by $\lVert f\rVert_\infty(h_n\mathcal H_n)^2$; and every term
selecting at least two non-identity factors is bounded in total by
$\lVert f\rVert_\infty\bigl(\prod_\ell(1+a_\ell)-1-\sum_\ell
a_\ell\bigr)$ with $a_\ell=2h_nH_{n,\ell}+(h_nH_{n,\ell})^2$, which is
at most an absolute constant times
$\lVert f\rVert_\infty(h_n\mathcal H_n)^2$ when
$h_n\mathcal H_n\le1$, and at most
$2\lVert f\rVert_\infty\le2\lVert f\rVert_\infty(h_n\mathcal H_n)^2$
otherwise, since all kernels involved are probability measures.
Taking the expectation over the independent per-position endpoint
samples replaces the sampled rates by the posterior-averaged rates in
the first-order term and gives
$|R_n|\le C\,h_n^2\,\E[\mathcal H_n^2]\,\lVert f\rVert_\infty$. \qed

\medskip
\noindent The bound depends on vocabulary size and sequence length
only through $\mathcal H_n$, which grows with $L$ and with
$\dot\beta(t)$. Since $\dot\beta$ diverges as $t\to1$, the first-order regime is the interior of the grid; at the final steps $h_n\mathcal H_n\gg1$ and the kernel is in effect a commitment to the sampled endpoint.

\section{Training recipe}
\label{app:training}

\begin{algorithm}[t]
\caption{Deployed sampling. The solve acts once, on the terminal
clean state; the controlled output is the decode of the corrected
continuous state, which is never re-quantised.}
\label{alg:inloop}
\begin{algorithmic}[1]
\Require text $c$, anchors $\mathcal A$, grid $t_0<\dots<t_{N-1}$,
solver budget $R$
\State $\vX_0 \gets$ uniform random tokens
\For{$n=0,\dots,N-2$}
  \State $\pi_n \gets$ anchor-conditioned guided clean-token
  posterior at $(\vX_n,c,\mathcal A)$
  \State $\vX_{n+1} \gets$ first event under rates sampled from
  $\pi_n$ \Comment{equations~\plaineqref{eq:endpoint_gen} and~\plaineqref{eq:first_event}}
\EndFor
\State $\pi_{N-1} \gets$ posterior at $(\vX_{N-1},c,\mathcal A)$;
\ \ $\vX_N \sim \pi_{N-1}$
\State $\vE_0 \gets \vC[\vX_N]$ \Comment{hard clean code embedding}
\State $\hat\vE \gets \operatorname{GN}(\vE_0;\mathcal A)$, $R$
accepted steps \Comment{\eqref{eq:gn_update}, identity metric}
\State \Return $\mathcal D(\hat\vE)$
\Comment{decode the corrected state directly}
\end{algorithmic}
\end{algorithm}

\noindent\textbf{Solver-coupled versus post-hoc training.} Both arms
are measured from the state their sampler emits before any solve,
under the shared solver of Table~\ref{tab:crossmodel} (validation,
$n{=}1024$). The solver-coupled sampler emits $0.314$\,m against
$0.367$\,m from the same model trained post hoc, whose sampler
carries no solver. Two accepted steps take them to $0.083$\,m, the deployed output,
and to $0.107$\,m; a centimetre takes $4.36$ accepted
steps against $4.56$, moving the state $0.338$ against $0.359$ in
relative norm. Mean contact speed runs $0.165$ against $0.183$\,m/s
at the emitted state and $0.201$ against $0.220$ after two steps.
Training through the solve lowers the emitted error from $0.367$ to
$0.314$\,m and keeps that margin along the solve.

Stage~1 trains the text-conditioned prior: sample
$t\sim\mathcal U(0,1)$, corrupt the clean token sequence with
$q_t(\cdot\mid z^\star)$ from \eqref{eq:path}, and train the
transformer to predict $z^\star$ with cross-entropy, dropping the text
condition with a fixed probability so that the same network supplies
the unconditional branch used by classifier-free guidance. Stage~2
adds the anchor conditioning of \S\ref{sec:model} and fine-tunes the
full model, still with cross-entropy alone, over a ladder of anchor
densities in which the first and last frames are always observed.
Stage~3 is the on-policy calibration. The model generates with its own sampler; anchor conditioning switches on at step $10$ of the $20$, the last four transitions retain gradients, and the terminal solve runs on the final state inside the graph through \eqref{eq:st} at $R{=}2$ accepted steps, $\lambda_0{=}10^{-2}$, four trials, a tolerance of $10^{-5}$ on the mean squared residual and the $48$-anchor row cap of the deployed configuration. The objective is
\begin{equation}
  \mathcal L = \mathcal L_{\mathrm{CE}} + 0.1\,\mathcal L_{\mathrm{local}},
  \qquad
  \mathcal L_{\mathrm{local}} = \overline{\lVert\mathcal D(\vE_{\mathrm{ST}})-\vY\rVert}_{\mathcal A}
  + 0.25\,\overline{\lVert\mathcal D(\vE_{\mathrm{ST}})-\vY\rVert}_{\mathcal H},
  \label{eq:stage3}
\end{equation}
with $\mathcal L_{\mathrm{CE}}$ the rollout cross-entropy on the sampler's own logits, and $\mathcal L_{\mathrm{local}}$ the mean Euclidean joint error of the straight-through decode against the reference clip, averaged over the anchored frame--joint pairs $\mathcal A$ and over the halo $\mathcal H$ of frames within two of an anchor. The solved state enters the objective through $\mathcal D(\vE_{\mathrm{ST}})$, whose forward value is the corrected state and whose gradient reaches the differentiable posterior state of \eqref{eq:st}. Stage~3 is short: the deployed checkpoint is taken at epoch $25$. The
path constants of \S\ref{sec:model} are $p{=}2$, $\alpha{=}0.9$ and
$c_{\beta}{=}3.0$, with $N{=}20$ sampler steps on a uniform grid over $[0.1,0.9]$, the tokens drawn uniformly at random at its first point, so $\dot\beta$ is evaluated away from both endpoints of the path, guidance scale $4.0$
and top-$k$ truncation to the highest-scoring tenth of the codebook.

\noindent\textbf{Sizes.} The masked transformer is $8$ layers of width
$384$ with $6$ heads and a $1536$-wide feed-forward block, dropout
$0.2$, and a conditioning drop probability of $0.1$ for
classifier-free guidance. The frozen tokeniser is a single quantiser
over $512$ codes of dimension $512$; its convolutional decoder is $512$ wide and $3$ deep with dilation growth $3$, and runs at unit temporal stride, neither downsampling nor upsampling in time. The residual encoder reads $54$ control
features per frame and injects through rank-$64$ adapters. Clips run
to $196$ frames. Training uses batch $48$ at learning rate
$5{\times}10^{-5}$, stepped by $0.1$ at $70$, $80$ and $90$ percent of
the schedule.

\bibliography{main}
\bibliographystyle{preprint}

\end{document}